\documentclass[aps,pra,twocolumn,superscriptaddress,showpacs,]{revtex4-1}

\usepackage{amsmath}
\usepackage{amssymb}
\usepackage{graphics}
\usepackage{graphicx}
\usepackage{epstopdf}
\usepackage{dcolumn}
\usepackage{bm}
\usepackage{xcolor}

\newcommand{\beq}{\begin{equation}}
\newcommand{\eeq}{\end{equation}}
\newcommand{\bary}{\begin{array}}
\newcommand{\eary}{\end{array}}
\newcommand{\beqary}{\begin{eqnarray}}
\newcommand{\eeqary}{\end{eqnarray}}
\newcommand{\lgl}{{\langle}}
\newcommand{\rgl}{{\rangle}}

\newcommand{\mtP}{\mathcal{P}}

\newcommand{\half}{\textstyle \frac{1}{2}}%
\newcommand{\ihalf}{\textstyle \frac{i}{2}}%
\newcommand{\quat}{\textstyle \frac{1}{4}}%
\begin{document}

\title{Rogue waves, self-similar statistics and self-similar intermediate asymptotics}

\author{Chunhao Liang}
\affiliation{Shandong Provincial Engineering and Technical Center of Light Manipulations \& Shandong Provincial Key Laboratory of Optics and Photonic Device, School of Physics and Electronics, Shandong Normal University, Jinan 250014, China}
\affiliation{Department of Electrical and Computer Engineering, Dalhousie University, Halifax, Nova Scotia, B3J 2X4, Canada}
\author{Sergey A. Ponomarenko}
\email[]{serpo@dal.ca}
\affiliation{Department of Electrical and Computer Engineering, Dalhousie University, Halifax, Nova Scotia, B3J 2X4, Canada}
\affiliation{Department of Physics and Atmospheric Science, Dalhousie University, Halifax, Nova Scotia, B3H 4R2, Canada}
\author{Fei Wang}
\affiliation{School of Physical Science and Technology, Soochow University, Suzhou 215006, China}
\author{Yangjian Cai$^{1,}$}
\affiliation{School of Physical Science and Technology, Soochow University, Suzhou 215006, China}

\date{\today}

\begin{abstract}
We advance a statistical theory of extreme event emergence in random nonlinear wave systems with self-similar intermediate asymptotics. We show, within the framework of a generic $(1+1)$D nonlinear Schr\"{o}dinger equation with linear gain, that extreme events and even rogue waves in weakly nonlinear, statistical open systems emerge as parabolic-shape giant fluctuations in the self-similar asymptotic propagation regime. We analytically demonstrate the self-similar structure of the non-Gaussian statistics of emergent rogue waves and validate our results with numerical simulations. Our results shed new light on generic statistical features of extreme events in nonlinear open systems with self-similar intermediate asymptotics.

\end{abstract}
\maketitle
\section{Introduction}
Rogue waves (RW), extremely rare, giant-amplitude waves obeying non-Gaussian statistics, were originally discussed in the oceanographic context~\cite{Ono,Zakh,RW-rev}. The concept has been quickly recognized as germane to generic wave supporting physics settings and RWs have been discovered, among others, in supercontinuum generating optical fibers~\cite{Solli,SCdud1}, optical cavities~\cite{Res1}, Bose-Einstein condensates~\cite{Blud}, Raman fiber amplifiers~\cite{Ham1,Ham2}, fiber lasers~\cite{RWlas1,RWlas2}, laser filamentation~\cite{Kasp}, plasmas~\cite{Mos}, stimulated Raman scattering~\cite{Yash1,Yash2}, discrete nonlinear lattices~\cite{latt}, and even in the multimode optical fibers and microwave transport in the linear propagation regime~\cite{RWf,Hell}.

Although there apparently exists no universal mechanism describing RW generation in any physical system~\cite{Dud-rev}, many weakly nonlinear and dispersive statistical wave systems are governed by a generic $(1+1)$D nonlinear Schr\"{o}dinger equation (NLSE)~\cite{Agra}. The RW excitation in the NLSE model with random input wave fields has been studied in the anomalous dispersion regime both numerically~\cite{Pic,Dud,Aga,Akh1, Akh2,Sur1,Sur2} and experimentally~\cite{Sur1,Sur2}. These studies revealed modulation instability driven RW excitation scenarios in weakly nonlinear conservative systems and elucidated the respective roles of  spontaneous Peregrine-like breather excitation from a noisy environment and of random soliton collisions in triggering the emergence of heavy-tailed probability density distributions (PDF) of wave intensities. Such heavy-tailed PDFs herald the RW generation in the system~\cite{Dud,Sur1,Sur2,Akh1,Akh2}.

At the same time, open physical systems often cannot support either solitons or breathers, at least as their long-term asymptotic states, because the energy supply from---or loss to---the environment precludes the establishment of precise balance between the dispersion and nonlinearity necessary for soliton formation. Although dissipative soliton formation is possible in some open systems if, on the one hand, the nonlinearity is balanced by dispersion/diffraction and, on the other hand, gain is balanced by loss~\cite{Dissol}, a multitude of open systems across physics disciplines exhibit self-similar dynamics instead. Examples range from blast waves in gas dynamics and turbulent bursts in fluids~\cite{Bar} to nonlinear waves in optical fiber~\cite{Krug1,Krug2,Ser} and graded-index waveguide~\cite{PSA2,PSA3} amplifiers, saturable two-level absorbers~\cite{PSA4}, and growing Bose-Einstein condensates~\cite{Drum,Vog}. Moreover, there exists a wide class of open wave systems displaying self-similar evolution in the intermediate range of parameters such that particular initial conditions at the source no longer play any role, though the system has not yet reached its steady-state~\cite{Bar,Krug1,Krug2,PSA4,Drum}. This observation prompts a natural question: Is there an universal scenario of extreme event generation in the statistical wave systems with self-similar intermediate asymptotics? A related fundamental issue has to do with the influence of self-similar dynamics on the wave ensemble statistics in the self-similar evolution regime.

In this work, we take the first step toward addressing these fundamental topics by advancing a statistical theory of extreme events in weakly nonlinear random wave systems with unsaturated gain, described within the framework of a generic NLSE modified by a linear gain term. The modified NLSE possesses self-similar intermediate asymptotics with a parabolic intensity profile in the {\it normal dispersion} regime~\cite{Krug2}. We develop a statistical theory of RW generation in the system by studying random input wave propagation there.  We analytically derive and numerically verify the PDF of a wave peak power ensemble, establish its non-Gaussian statistics, and demonstrate the self-similar evolution of the ensemble statistics on random pulse propagation in the intermediate regime. We stress that,  to our knowledge, this is the first demonstration of {\it the RW statistics self-similarity} in any nonlinear random wave system. Our analytical results are independent of a particular source ensemble model. In addition, they are in excellent agreement with numerical simulations of the modified NLSE.

As the NLSE with linear gain captures salient features of optical wave propagation in any weakly nonlinear amplifying media~\cite{Krug1,Krug2}, our findings are expected to be generic. We also note that since the NLSE with a linear gain term and harmonic trapping potential attains a self-similar asymptotics as well~\cite{Drum}, our results apply, at least qualitatively, to optical waves in graded-index waveguide amplifiers and matter waves in growing Bose-Einstein condensates.

This work is organized as follows. In Section I, we introduce a generic dimensionless NLSE with linear gain as our mathematical model of self-similar dynamics in a wide variety of nonlinear wave systems and briefly review its parabolic self-similar solutions. In Section II, we formulate our statistical source ensemble model and present numerical results for a typical ensemble member evolution. In Section III, we present an analytical derivation of the wave peak power PDF in the self-similar regime and verify our findings with comprehensive numerical simulations of the modified NLSE. We present our conclusions in Section IV.

\section{ Modified dimensionless nonlinear Schr\"{o}dinger equation and its parabolic self-similar solutions}
We consider  statistical pulse propagation in a  nonlinear fiber amplifier in the normal dispersion regime governed by the modified NLSE in the form
\beq\label{GNLSE-dim}
		i\partial_{Z}U -\half\beta_2\partial_{TT}^2 U +\gamma |U|^2 U -\ihalf GU=0,
			\eeq
where $G$ is linear gain; $\beta_2$ is a group-velocity dispersion, and $\gamma$ is a Kerr nonlinearity coefficients. To explore generic features of random waves with self-similar intermediate asymptotics, which are  independent of the source and medium particulars, it will prove convenient to work with {\it dimensionless} variables, defined as
	\beq
		\Psi =U/\sqrt{\lgl P_0 \rgl}; \hspace{0.5cm} t=T/T_p, \hspace{0.5cm} z=Z/L.
			\eeq
Here $T_p$ is an average pulse width and $\lgl P_0\rgl $ is an average peak power of a statistical input pulse ensemble; $L=\sqrt{L_{NL}L_D}$, where $L_{NL}=(\gamma \lgl P_0\rgl )^{-1}$ and $L_D =T_p^2/\beta_2$ are the usual nonlinear and dispersion lengths.  In the dimensionless variables, the modified NLSE reads
	\beq\label{GNLSE-ndim}
		i\sigma\partial_{z}\Psi -\half\sigma^2\partial_{tt}^2 \Psi + |\Psi|^2 \Psi -\ihalf \sigma g\Psi=0,
			\eeq
where we introduced a dimensionless gain $g=GL$ and soliton $\sigma=\sqrt{L_{\mathrm{NL}}/L_{\mathrm{D}}}$ parameters which entirely determine the system dynamics given the source coherence state. We note that the smaller the soliton parameter, the faster the self-similarity is attained~\cite{Krug1}.

We now present a parabolic self-similar solution to Eq.~(\ref{GNLSE-ndim}) by transforming the original results of~\cite{Krug1}  to a dimensionless form.
To this end,  we express the parabolic self-similar solution (similariton) in the polar form
	\beq\label{Psi-SS}
		\Psi=Ae^{i\phi/\sigma},
			\eeq
where the amplitude and phase are given by the expressions
\beq\label{SSA}
		A(t,z)=a(z)f[t/t_p(z)],
			\eeq
and
	\beq\label{SSphi}
		\phi(t,z)=\phi_0(z)+c(z)t^2,
			\eeq
respectively. Henceforth we are mainly focusing on the similariton amplitude as we will derive the similariton peak power PDF; thus we ignore the phase. In Eq.~(\ref{SSA}),
$a(z)$ and $t_p(z)$ are dimensionless amplitude and width of the similariton, given by
	\beq\label{a}
		a(z)=a_0 e^{gz/3},
			\eeq
and
	  \beq\label{tp}
		t_p (z)=a_0^{-2}e^{gz/3},			\eeq
respectively.
Here
	\beq\label{a0}
		a_0=\half (\sqrt{2}g w_0)^{1/3},  \hspace{1cm} 	w_0=W_0/\lgl W_0\rgl,
			\eeq
where the dimensionless input pulse energy $w_0$ is normalized to the average energy of the input pulse ensemble,
	\beq\label{W_0}
		\lgl W_0\rgl =\int_{-\infty}^{\infty}dT\,\lgl |U(T,0)|^2 \rgl.
					\eeq
Here the angle brackets denote ensemble averaging. Further, introducing the dimensionless similarity variable, $\eta=t/t_p(z)$, we can express the similariton profile as
\beq\label{f}
		f(\eta)=\left\{\bary{cc}
				 \sqrt{1-\textstyle\frac{g^2}{18a_0^6}\eta^2},  & |\eta|\leq 3\sqrt{2}a_0^3/g, \\
				  0  , &  |\eta|\geq 3\sqrt{2}a_0^3/g.
				  	\eary
						\right.
                       \eeq
 It follows at once from Eqs.~(\ref{SSA}) through~(\ref{f}) that the similariton power profile has a parabolic shape which, together with its amplitude and width, is completely determined by the (scaled) input pulse energy
 $w_0$ as well as the medium gain $g$.  To study the individual pulse evolution numerically, we have to specify a random pulse ensemble at the source.

\section{Statistical ensemble formulation of input pulses}
We now describe the input pulse ensemble in terms of a generic Gaussian Schell model ~\cite{MW} previously employed in extreme event studies~\cite{Yash1,Yash2}. The(GSM) ensemble has a Gaussian average intensity and Gaussian degree of the second-order temporal coherence~\cite{MW, Laleh}. The GSM ensemble mutual intensity, defined as
\beq
	\Gamma_0(t_1, t_2)=\lgl \Psi^{\ast}(t_1,0)\Psi(t_2,0)\rgl,
		\eeq
reads then
	\beq
		\!\Gamma_0 (t_1, t_2)\!=\!\frac{1}{\sqrt{\pi}}\exp\left(-\frac{t_1^2+t_2^2}{2}\right)\exp\left[-\frac{(t_1-t_2)^2}{2\sigma_c^2}\right].\!\!
			\eeq
Here $\sigma_c=T_c/T_p$ is a source coherence parameter. Given $\sigma_c$, we can approximate any realistic GSM source with a finite number $N=N(\sigma_c)$ of (uncorrelated) excited coherent modes
via the Karhunen-Lo\`{e}ve expansion,
\beq
	\Psi(t,0)=\sum_{n=0}^{N} c_n \psi_n(t),
		\eeq
where $\{c_n\}$ are complex random amplitudes and $\{\psi_n(t)\}$ are coherent mode functions, known for a GSM source to be~\cite{MW}
\beq
 \psi_{n}(t)=\left(\frac{2\xi}{\pi}\right)^{1/4}\left(\frac{1}{2^{n}n!}\right)^{1/2}\,H_{n}(\sqrt{2\xi}t)e^{-\xi t^2}.
			\eeq
Here $H_{n}(x)$ is a Hermite polynomial of the order $n$, and	 we introduced the notations
	\beq\label{alf}
		\alpha=1/2,  \hspace{0.5cm} \beta=1/(2 \sigma_{c}^2),
			\eeq
and
	\beq\label{xi}
		\xi=\sqrt{\alpha^2 + 2\alpha\beta}.
			\eeq
We note that the mode functions are orthonormal such that
	\beq
		\int_{-\infty}^{\infty}dt\,\psi_m^{\ast}(t)\psi_n(t)=\delta_{mn}.
			\eeq			
The second-order statistics of the (uncorrelated) random amplitudes $\{ c_n \}$ are determined by
	\beq
		\lgl c_n^{\ast} c_m\rgl =\lambda_n \delta_{mn},
			\eeq
with the modal weight distribution,
	\beq
		\lambda_n = A\left( \frac{\beta}{\alpha +\beta+\xi}\right)^n.
		\eeq
		
The set of $\{\lambda_n\}$ determines average energies of the excited coherent modes representing the source. The constant $A$ is determined by a relevant physical normalization; in our case, the (scaled) average energy $\lgl w_0 \rgl$ of the pulse must be normalized to unity:
	\beq
		1=\int_{-\infty}^{\infty} dt\,\lgl |\Psi(t,0)|^2\rgl =\sum_{n=0}^{N}\lgl |c_n|^2\rgl =\sum_{n=0}^N \lambda_n,
					\eeq		
implying that
	\beq
		\lambda_n=\left( \frac{\beta}{\alpha +\beta+\xi}\right)^n \frac{1-\frac{\beta}{\alpha+\beta+\xi}}{1-\left(\frac{\beta}{\alpha+\beta+\xi}\right)^{N+1}}.
			\eeq
Note, the case $N=0$ implies there are no excited modes, yielding $\lambda_n=\lambda_0=1$ as expected; that is, only a ground mode of the source is excited---this is an example of a fully (temporarily) coherent source at the second order.		

To specify the source ensemble beyond the second order, we express the set of $\{c_n\}$ in the polar form
	\beq
		c_n=\sqrt{E_n}e^{i\phi_n},
			\eeq
and assume the following joint PDF of the phases and energies
	\beq\label{PDF0}
		\mtP(E_n,\phi_n)=\frac{\theta(E_n)}{2\pi\lambda_n}e^{-E_n/\lambda_n}.
			\eeq
 Here $\theta(x)$ is a Heaviside step function; in other words, the phases are uniformly distributed in the interval, $-\pi\leq\phi_n\leq \pi$, and the mode powers obey the Rayleigh distribution. Eq.~(\ref{PDF0}) guarantees  Gaussian statistics of the input pulse ensemble $\{\Psi(0,t)\}$. We stress that coherent mode amplitude fluctuations are crucial as they determine the total source energy fluctuations, which, in turn, shape the RW statistics in the self-similar intermediate propagation regime.

We performed numerical simulations of a GSM ensemble of  $10^4$ random pulses. In our numerical simulations we take $T_p=0.2$ ps, $\lgl P_0 \rgl=2.8$ W, $\gamma=5.8$ W$^{-1}$ m$^{-1}$, $\beta_2=0.025$ ps$^2$m$^{-1}$, and $G=1.9$m$^{-1}$, implying that the system is in the nonlinearity dominated regime, $\sigma\ll1$ corresponding to the experiment~\cite{Krug1}, except the average input pulse ensemble power and fiber nonlinearity are scaled by three orders of magnitude down and up, respectively. Thus, the nonlinear length remains the same and the parabolic self-similar regime is reached at sufficiently short amplifier lengths. The reason behind the scaling is practical difficulty to generate high-power (kW) ultrashort pulse sources with thermal-like power fluctuations, whereas engineering such a statistical source with the power around  1 W has been recently reported~\cite{Sur1}. On the other hand, fiber amplifiers with such high nonlinearities can, in principle, be realized, for instance, by dye doping high nonlinear refractive index liquids, filling the cores of specially designed photonic crystal fibers of the kind reported in c.f.~\cite{Giess,Yash3}. Alternatively, one can attain the self-similar regime with statistical beams propagating in highly nonlinear chalcogenide planar waveguide amplifiers with defocusing nonlinearities of the order of $\gamma\simeq 10$ W$^{-1}$m$^{-1}$~\cite{Eggl}.

\begin{figure}[t]
\centering
   \includegraphics[width=\linewidth]{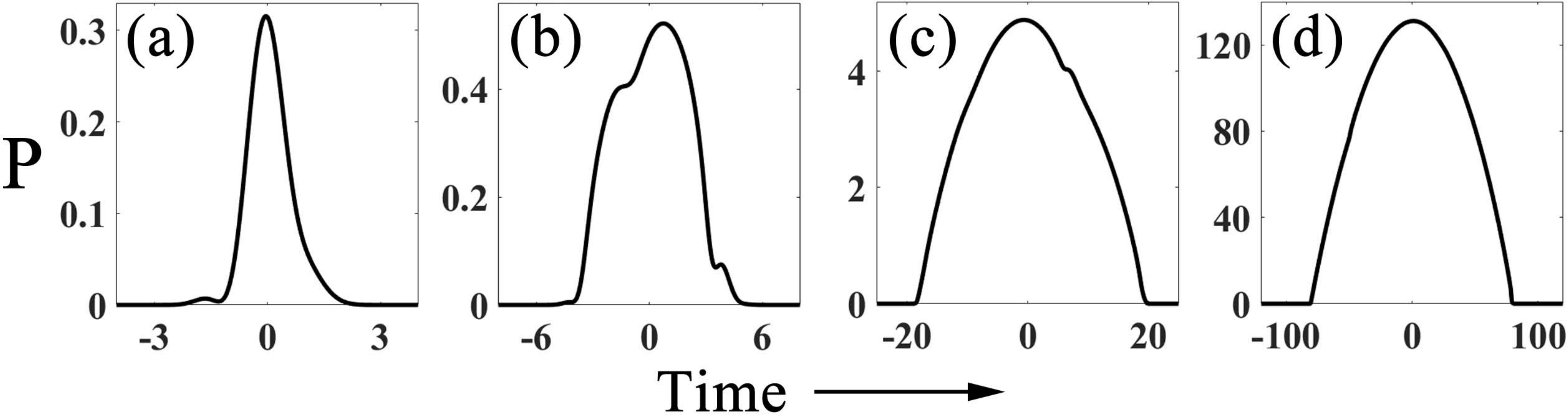}
   \caption{Power profile of a random pulse ensemble realization as a function of the dimensionless time $t=T/T_p$ at (a) $Z=0$, (b) $Z=1$ m, (c) $Z=3$ m, and (d) $Z=5.5$ m. The dimensionless power is $P=|U|^2/\lgl P_0\rgl$  where $\lgl P_0\rgl =\lgl W_0\rgl/\sqrt{\pi} T_p$ is an average peak power of the GSM ensemble expressed in terms of its average energy $\lgl W_0\rgl $ and the pulse width $T_p$. The source coherence parameter is taken to be $\sigma_c=1$. }
   \label{figure1}
\end{figure}
A particular ensemble realization evolution is illustrated in Fig.1. We can infer from the figure that the power profile of the realization attains a parabolic shape at the distance $Z=5.5$m; this conclusion holds for any realization, albeit the propagation distance over which the parabolic profile is reached varies from realization to realization. Most importantly, as soon as the parabolic profile is reached, it remains unchanged, up to scaling, indicating the self-similar regime ensues.

\section{Self-similar statistics of peak power PDF of the pulse ensemble}
The self-similar dynamics of each realization results in a remarkably simple statistical evolution of the ensemble as a whole which can be revealed by analytically deriving its peak power PDF. To this end, we can find the total energy distribution of the source pulse ensemble. As the system dynamics is entirely determined by the source energy $w_0$, we derive a generic source energy PDF without committing to a specific source model under the only assumptions that (i) the mode powers have thermal-like distributions of Eq.~(\ref{PDF0}) and (ii) we know the set of average mode energies $\{\lambda_n\}$. The source energy is then given by the expression
 	\beq
		w_0=\int_{-\infty}^{\infty}dt |\Psi(t,0)|^2 =\sum_{n=0}^N |c_n|^2 =\sum_{n=0}^N E_n.
			\eeq
As all modes are uncorrelated, we can use characteristic functions to determine $\mtP(w_0)$. To this end, we first determine a characteristic function of $E_n$,
	\beq
		\chi_{\rho_n }(s)=\lgl e^{is E_n}\rgl=\frac{1}{1-is\lambda_n}.
			\eeq
It follows at once that
	\beq\label{chi-w0}
		\chi_{w_0 }(s)=\prod_{n=0}^N  \frac{1}{1-is\lambda_n}.
			\eeq
Taking an inverse Fourier transform of Eq.~(\ref{chi-w0}), we obtain the source energy PDF as
	\beq\label{aux1}
		\mtP(w_0)=\int_{-\infty}^{\infty}\frac{ds}{2\pi} e^{-isw_0}\prod_{n=0}^N  \frac{1}{1-is\lambda_n}.
			\eeq
Eq.~(\ref{aux1}) can be cast into the form
	\beq\label{PDFw0}
		\mtP(w_0)=\int_{-\infty}^{\infty}\frac{ds}{2\pi} e^{-isw_0}\prod_{n=0}^N  \frac{i/\lambda_n}{s+i/\lambda_n}.
			\eeq
Assuming the source modes have no degeneracy, which is  certainly true of the GSM source modes, all $\lambda's$ are distinct. It follows that the integral on the r.h.s of Eq.~(\ref{PDFw0}) is straightforward to do in the complex plane---the result is a sum of residues at simple poles,
$s_m=-i/\lambda_m$, $0\leq m\leq N$. The result then reads
	\beq
		\mtP(w_0)=\sum_{m=0}^{N}\frac{1}{\lambda_m}e^{-w_0/\lambda_m}\prod_{ \underset{n\neq m}{n=0}
		                                                                                 }^{N} \frac{\lambda_n^{-1}}{\lambda_n^{-1}-\lambda_m^{-1}}.
		                                                                                                                                       \eeq
Note the source energy PDF is a weighted superposition of exponential distributions of the modes carrying the average energies $\{\lambda_n \}$.
		
Next, let us write down the pulse peak power as
 	\beq\label{Peak}
		P_\ast (w_0, z)=|\Psi(0,z)|^2=\quat (\sqrt{2}gw_0)^{2/3} e^{2gz/3},
			\eeq
where we made use of Eqs.~(\ref{Psi-SS}) through~(\ref{a0}). The peak power PDF of the pulse ensemble is defined as
	\beq\label{PDF}
		\mtP(P_\ast, z)=\lgl \delta[P_\ast -P_\ast(w_0, z)]\rgl_{w_0},
			\eeq
where the angle brackets denote averaging over an ensemble of random incident pulses. The averaging in Eq.~(\ref{PDF}) can be carried out using the delta function property,
	\beq
		\delta[P_\ast -P_\ast (w_0,z)]=\sum_{w_{0\pm}}\frac{1}{|\partial_{w_0}P(w_0,z)|}\delta(w_0-w_{0\pm}).
			\eeq
Here $w_{0\pm}$ are the two roots of the equation
	\beq
		P_\ast=P_\ast (w_0,z),
					\eeq
which are written explicitly  as
	\beq\label{Wpm}
		w_{0\pm}=\pm \frac{8e^{-gz}}{\sqrt{2}g}P_\ast^{3/2}.
			\eeq
However, only the positive root is physical because $w_0 \geq 0$. We can then drop the other root, perform a trivial integration with the delta function and, collecting all terms, we obtain a properly normalized PDF as
	\beq\label{SSPDF}
		\mtP(P_\ast, z)\!=\!e^{-gz}\sqrt{P_\ast}\sum_{m=0}^{N}b_m \exp\left(-\frac{8 e^{-gz}}{\sqrt{2}g\lambda_m}P_\ast^{3/2}\right),\!\!
		\eeq
where the set of normalization constants $\{ b_m\}$  reads
\beq\label{Norm}
	b_m=\left(\frac{12}{\sqrt{2}g\lambda_m}\right)\,\frac{\prod_{ \underset{n\neq m}{n=0}}^{N} \,\frac{\lambda_n^{-1}}{\lambda_n^{-1}-\lambda_m^{-1}}}{\sum_{m=0}^{N}\prod_{ \underset{n\neq m}{n=0}}^{N} 						\,\frac{\lambda_n^{-1}}{\lambda_n^{-1}-\lambda_m^{-1}}}.
		\eeq
Here $N$ coherent modes can correspond to physical modes of a multimode fluctuating source. The normalization constants are found from the condition $\int_0^{\infty} dP_\ast \mtP(P_\ast, z)=1$. Eq.~(\ref{PDF-SS}) happens to represent a weighted superposition of Weibull distributions with the index $3/2$~\cite{Wei}. Finally, we observe that the PDF can be expressed in a manifestly self-similar form as
	\beqary\label{PDF-SS}
		\mtP&&(P_\ast, z)=e^{-2gz/3}\sqrt{P_\ast e^{-2gz/3}} \nonumber \\
			&&\mbox{}\times\sum_{m=0}^{N}b_m \exp\left[-\frac{8 }{\sqrt{2}g\lambda_m}(P_\ast e^{-2gz/3})^{3/2}\right],	                                                                                                                                       \eeqary
in terms of the similarity variable $P_\ast e^{-2gz/3}$.  We display the PDF in Fig 2 (left panel) at three propagation distances: $Z=4.5$, $Z=5$, $Z=5.5$ meters in black solid, red dashed and blue dash-dotted curves, respectively. To visualize the self-similarity, we plot the PDF in the scaled variables, explicitly demonstrating in Fig. 2  (right panel) that all three curves coalesce into one. We stress that the derived {\it self-similar structure} of the PDF is independent of either a specific source model or the source coherence level, although the source coherence parameter $\sigma_c$ affects the overall PDF shape as we will show below. Thus, our fundings apply to any statistical pulse ensemble with thermal-like power fluctuations in the self-similar evolution regime.
\begin{figure}[t]
\centering
   \includegraphics[width=\linewidth]{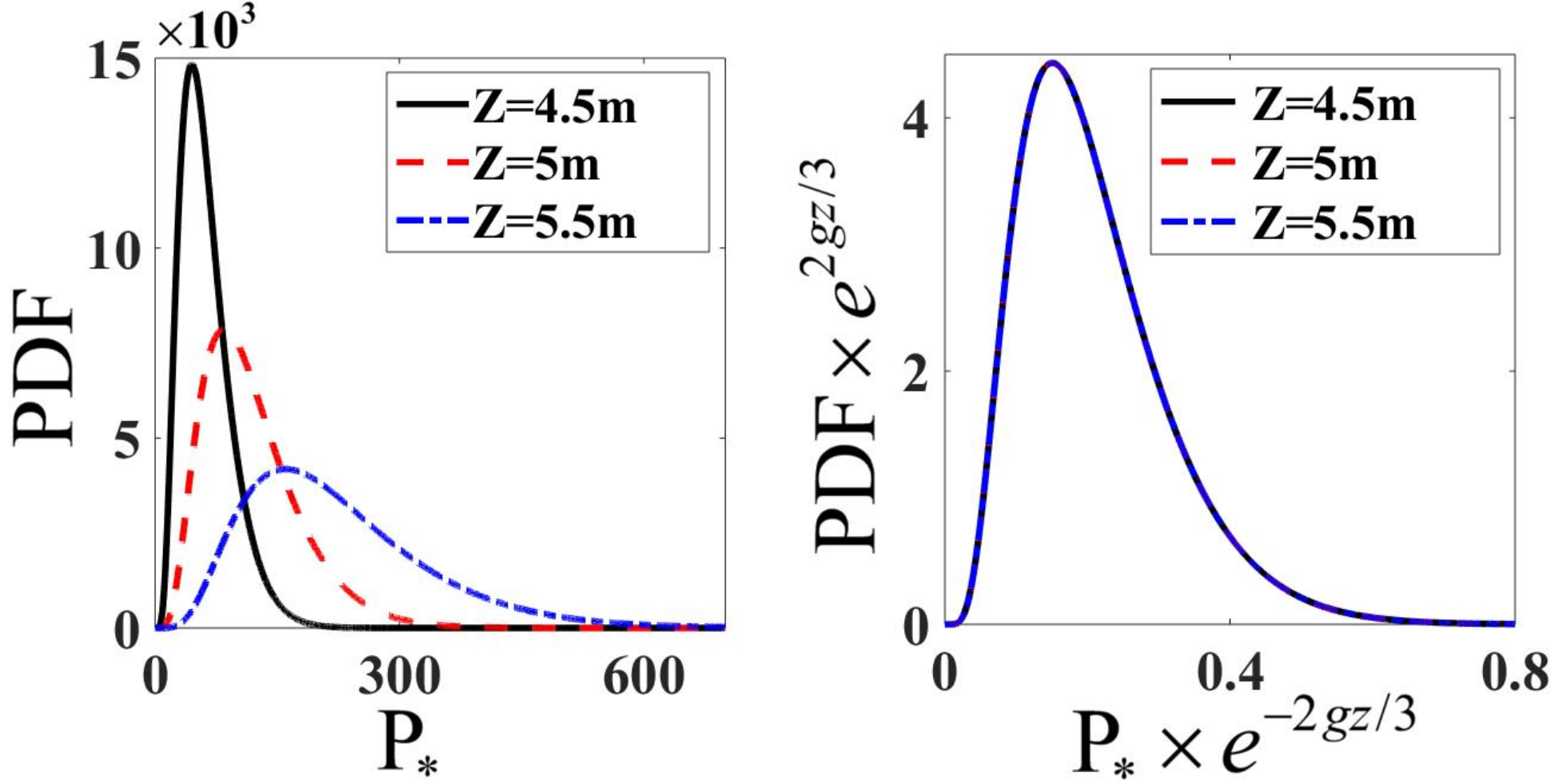}
   \caption{(Color online) Left: Analytical PDF of the dimensionless peak power $P_\ast$ of the pulse ensemble at three propagation distances: $Z=4.5$ m (black solid curve), $Z=5$ m (red dashed curve) and $Z=5.5$ m (blue dash-dotted curve). Right: PDF of the pulse peak power in the scaled variables $\mtP e^{2gz/3}$ and $P_\ast e^{-2gz/3}$ at the same propagation distances. The source coherence parameter is taken to be $\sigma_c=1$. }
   \label{figure2}
\end{figure}

To verify our analytical results and ascertain the RW existence among the extreme events in the self-similar intermediate propagation regime, we performed numerical simulations of the modified NLSE, Eq.~(\ref{GNLSE-ndim}). The results are shown in Fig. 3 for the same propagation distances as in Fig.2. As is evidenced by the figure, numerical PDFs, shown by green histograms, are in excellent agreement with the theoretical curves, confirming that the system is indeed in the self-similar regime at these propagation distances. The RW emergence is marked by the abnormality index (AI) greater than two~\cite{Dud-rev}. The AI is defined as the ratio of an RW intensity $I_{RW}$ to that of a ``significant intensity" $I_{1/3}$,
	\beq
		AI=I_{\mathrm{RW}}/I_{1/3},
			\eeq
where the ``significant intensity" is defined as the mean intensity of one third of the highest peak intensity events. We can see that at all selected distances $AI\geq 2$, indicating RW emergence in the system.  We stress that we limit ourselves to statistical signatures of RW emergence; the examination of the other characteristic of RWs as the waves appearing from nowhere and disappearing without a trace~\cite{Taki,Akh3}  lies outside the scope of this work. In the inset panels to Fig. 3 we exhibit the corresponding RWs in red curves which clearly acquire a parabolic shape. Moreover, we display the average intensity distribution of the pulse ensemble at the corresponding propagation distance in the same inset panel in a black dashed curve to show that it has a Gaussian-like rather than the parabolic shape. This is because each ensemble realization attains a parabolic shape with a different pulse width and peak pulse power, given by Eqs.~(\ref{tp}) and~(\ref{Peak}), respectively, such that the ensemble average over these parabolic pulse profiles yields a Gaussian-like pulse. Thus, we conclude that extreme events in general---and RWs, in particular---emerge as giant parabolic shape fluctuations away from the average in the self-similar intermediate asymptotic regime. Further, we note that the AI should remain the same for a given pulse ensemble in this regime because $I_{RW}$ and $I_{1/3}$ scale the same way with the propagation distance there. This observation is borne out by our numerical simulations displayed in Fig. 3: We can infer from the figure that $AI\simeq 2.4$ up to numerical round-off errors.
\begin{figure}[t]
\centering
   \includegraphics[width=\linewidth]{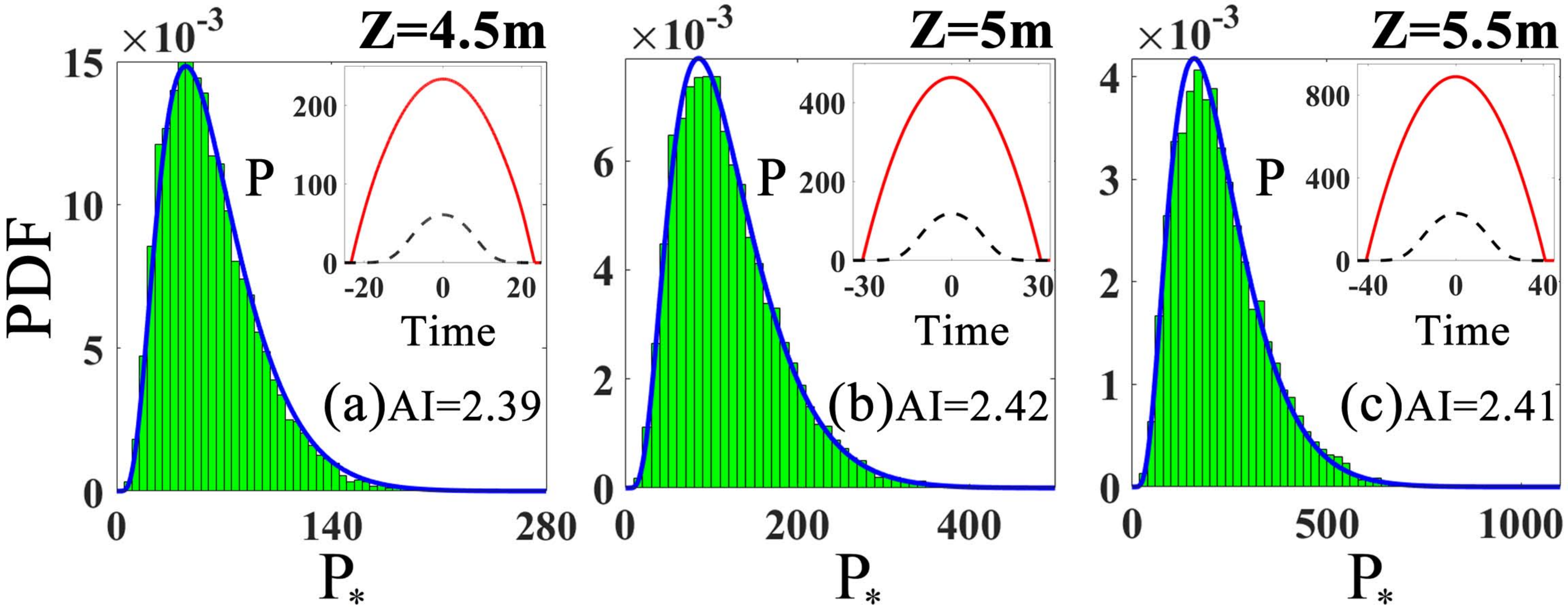}
   \caption{(Color online) Analytical (solid blue curve) and numerical (green histogram) PDF of the peak pulse power as a function of the dimensionless peak power $P_\ast$ at (a) $Z=4.5$ m, (b) $Z=5$m and (c) $Z=5.5$m. The source coherence parameter is taken to be $\sigma_c=1$. The abnormality index AI is also given. Insets: RW (solid red curve) and average pulse ensemble (dashed black curve) power profiles in the dimensionless units, $P=|U|^2/\lgl P_0\rgl$ and $t=T/T_p$, at the same propagation distances. }
   \label{figure3}
\end{figure}
\begin{figure}[t]
\centering
   \includegraphics[width=\linewidth]{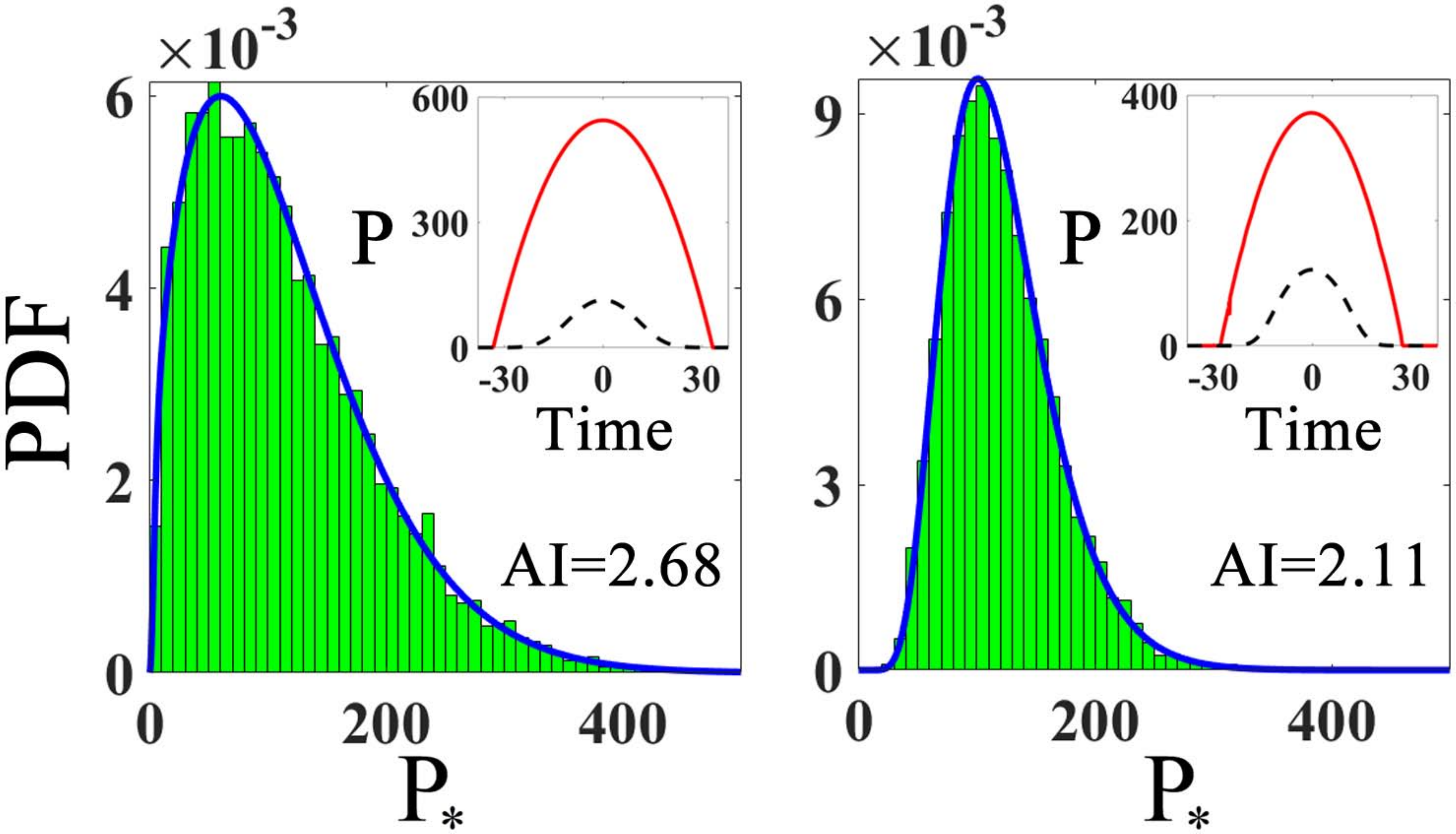}
   \caption{(Color online) Analytical (solid blue curve) and numerical (green histogram) PDF of the peak pulse power as a function of the dimensionless peak power $P_\ast$ at $Z=5$m for $\sigma_c=10$ (left) and $\sigma_c=0.5$ (right). The abnormality indices AI are also given: $AI=2.68$ (left), and $AI=2.11$ (right). Insets: RW (solid red curve) and average pulse ensemble (dashed black curve) power profiles in the dimensionless units, $P=|U|^2/\lgl P_0\rgl$ and $t=T/T_p$, at the same propagation distance. }
   \label{figure4}
\end{figure}

Finally, we examine the ensemble peak power PDF dependence on the source coherence state. To this end, we performed simulations for a very coherent, $\sigma_c=10$, and rather less coherent  $\sigma_c=0.5$ pulse ensembles and compared their PDFs in the self-similar asymptotic propagation regime in Fig. 4 using Eqs.~(\ref{SSPDF}) and~(\ref{Norm}) as well as our numerical data. We can infer from Fig. 4 by comparing either analytical PDF curves or the histograms on the left and right panels that as the source coherence increases---and so does $\sigma_c$---the PDF tail stretches as well. Indeed,  $AI\simeq 3$ for the very coherent source, while $AI\simeq 2$ for the less coherent one in Fig. 4. This conclusion appears to be at odds with our previous results on RW generation in stimulated Raman scattering with a noisy Stokes pulse input ensemble~\cite{Yash2}. To reconcile the two observations, we recall that as $\sigma_c$ increases, so does the effective number of excited coherent modes of the source~\cite{MW}. Further, the Raman nonlinearity has very long memory implying that the greater the number of coherent modes, the greater the chances for a (giant power) ``champion" mode to emerge within a Stokes pulse ensemble. This is because Raman medium memory reinforces unequal energy redistribution from a pump pulse among the Stokes pulse coherent modes throughout multiple Raman scattering cycles. In the case of weakly nonlinear amplifying media, however, the instantaneous Kerr nonlinearity has no coherent memory. Consequently, the lack of cumulative reinforcement of unequal power gain among the pulse ensemble modes favours the chances of giant power mode emergence for sources with a few excited coherent modes.

\section{Conclusions}
In conclusion, we have elucidated the emergence scenario and salient statistical properties of extreme events in weakly nonlinear, statistical open systems exhibiting self-similar intermediate asymptotic evolution. We have demonstrated that rogue waves manifest themselves as giant self-similar fluctuations away from the average and they acquire self-similar, non-Gaussian statistics in such systems. We stress that our generic results hold in the intermediate evolution regime where gain saturation and ensuing amplified spontaneous emission noise are negligible. As the pulse intensity will have sufficiently grown up, we can no longer neglect photon emission by the excited levels of the medium atoms, stimulated by the pulse field; the stimulated emission, in turn, causes gain saturation. At the same time, spontaneously emitted photons stimulate the emission of more random photons, adding amplified spontaneous emission noise to the system. The latter, which can be the dominant noise contribution at high pulse amplification levels, is expected to ultimately lead to the break-down of the system self-similarity, cause the eventual destruction of statistically self-similar RWs and hence invalidate the proposed generic RW excitation mechanism beyond the intermediate evolution regime. The RW nature and statistical properties near gain saturation in presence of pronounced amplified spontaneous emission is a challenging open problem which we plan to address in the future.

S.A.P. acknowledges financial support from Natural Science and Engineering Research Council of Canada, (RGPIN-2018-05497); F.W. acknowledges financial support from National Natural Science Foundation of China, (11874046); Y.C. acknowledges financial support from National Natural Science Foundation of China, (91750201, 11525418).

\end{document}